\documentstyle[multicol,prl,aps,epsf,psfig]{revtex}

\begin{document}

\newcommand{\be}{\begin{equation}}
\newcommand{\ba}{\begin{eqnarray}}
\newcommand{\ee}{\end{equation}}
\newcommand{\ea}{\end{eqnarray}}
\newcommand{\non}{\nonumber}
\newcommand{\ct}{\cite}
\newcommand{\bi}{\bibitem}
\title
{Coulomb gap in one-dimensional disordered electronic systems}

\author{Amit Dutta$^1$, Lars Fritz$^2$ and Diptiman Sen$^3$}
\address{$^1$ Department of Physics, Indian Institute of Technology,
Kanpur 208016, India \\
$^2$ Institut f\"ur Theoretische Physik und Astrophysik,
Universit\"at W\"urzburg, Am Hubland, 97074 W\"urzburg, Germany \\
$^3$ Centre for Theoretical Studies, Indian Institute of Science,
Bangalore 560012, India \\}
\maketitle
 
\begin{abstract}

We study a one-dimensional system of spinless electrons in the presence of a 
long-range Coulomb interaction (LRCI) and a random chemical potential at each 
site. We first present a Tomonaga-Luttinger liquid (TLL) description of the 
system. We use the bosonization technique followed by the replica trick to 
average over the quenched randomness. An expression for the localization 
length of the system is then obtained using the renormalization group method 
and also a physical argument. We then find the density of states
for different values of the energy; we get different expressions 
depending on whether the energy is larger than or smaller than the inverse
of the localization length. We work in the limit of weak disorder where the 
localization length is very large; at that length scale, the LRCI has the 
effect of reducing the interaction parameter $K$ of the TLL to a value much 
smaller than the noninteracting value of unity. 
\vskip .4 true cm

PACS number(s): ~71.10.Pm, ~71.27.+a, ~71.20.-b

\end{abstract}

\begin{multicols}{2}

\section{Introduction}

In a strongly localized electron system with an unscreened long-range Coulomb 
interaction (LRCI) between the electrons, the density of states (DOS) is 
expected to exhibit a gap at the Fermi energy. This is called the ``Coulomb 
gap" \ct{efros} as it arises due to the LRCI. A phenomenological argument 
\ct{efros} suggests that the DOS should obey a power-law of the form
\be
D(\omega) \sim |\omega|^{d-1} ~,
\ee
where $d$ is the spatial dimensionality of the system, and $\omega$
is measured from the Fermi energy. We shall consider here the one-dimensional
case where the DOS is expected to show a logarithmic behavior in the
extremely localized limit \ct{white,raich,mobius,vojta}.

Following the recent developments in carbon nanotube technology 
\ct{ebbesen}, there has been an upsurge in the studies of one and quasi-one 
dimensional electron systems. In low dimensional systems, the 
electron-electron interactions play a dominant role leading to a behavior 
significantly different from that of conventional Fermi liquids. Short-range
repulsive interactions between the electrons lead to the Tomonaga-Luttinger
liquid (TLL) behavior \ct {voit,schulzcp,sen}, whereas the LRCI is believed
to lead to a Wigner crystal \ct{schulz}.

It is well known that for a one-dimensional noninteracting electron system with
random disorder, all the states are localized due to repeated back-scatterings
of the electrons \ct{leer}. However, the DOS of noninteracting electrons is 
finite at the Fermi energy \ct{halperin}. The presence of LRCI together with 
random impurities modifies the DOS in a drastic way and supposedly
leads to the Coulomb gap behavior as discussed above. In an earlier study, 
Vojta and John \ct{vojta} studied a one-dimensional electron system, with 
LRCI and randomness, in the extremely localized limit where the overlap of
electronic wave functions can be neglected. They find the form of the DOS to be
\ct{vojta}
\be
D(\omega)\sim \left( \ln \frac {E_c}{|\omega|} \right)^{-1} ~,
\label{dos0}
\ee
where $E_c$ is a cut-off energy \ct{raich,mobius}. In a recent paper, Lee
\ct{lee} derived the expression of the DOS of a one-dimensional system of 
spinless electrons with LRCI and impurities at random positions in the opposite
limit where the quantum effects (electron hopping) play a dominant role, 
and the pinning is weak, i.e., the localization length is much larger 
than the inter-impurity distance. Following studies of a pinned Wigner crystal
\ct{maurey,fukuyama} and using a semiclassical approximation, Lee \ct{lee} 
finds the DOS at low energy to be
\be
D(\omega) \sim |\omega|^{\sqrt{1+ \eta}/2} ~,
\label{dos1}
\ee
where the exponent $\eta$ (defined in Eq. (\ref{eta}) below) is determined by 
the localization length of the system. This power-law behavior is consistent 
with the existing numerical studies \ct{jeon}.

In this paper, we study a model of spinless electrons with LRCI as in 
Ref. \ct{lee} (with quantum effects, i.e., with the electron hopping term)
in the presence of a random chemical
potential at each site. In our approach, we find that a TLL picture \ct{sen},
rather than a semiclassical approach \ct{lee}, provides a convenient
description of the system. The localization length $L_0$
is derived using a renormalization group (RG) study of the effective 
disorder-averaged bosonized action \ct{giamarchi}, and is found to be very 
large in the limit of weak disorder. (We also present a simple physical
understanding of the expression for $L_0$). Due to the LRCI, the interaction 
parameter $K$ of the TLL is found to be effectively a function of the length
scale. At the length scale $L_0$, the value of $K$ is given by
a value much smaller than unity (which is the value of $K$ 
for the noninteracting system). We should remark here that the present 
one-dimensional disordered electron system with LRCI appears to be an unique 
example of a TLL with $K << 1$. 

Once we make use of a TLL description, the DOS can be found in a standard way 
\ct{sen,gogolin}. We will show that for $\omega \gtrsim v_F /L_0$, the DOS 
is given by 
\ba
D(\omega) & \sim & |\omega|^\beta ~, \non \\
\beta &=& \frac{{\sqrt {1+\eta}}}{2} + \frac{1}{2{\sqrt {1+\eta}}} - 1 ~,
\label{dos2}
\ea
where, {\it unlike} a standard TLL, $\eta$ itself depends 
on $\omega$ as discussed below. Eq. (\ref{dos2})
resembles the result of Lee \ct{lee} in the limit $\eta >>1$; however, 
$\eta$ is not taken to be a function of $\omega$ in Ref. \ct{lee}. On the
other hand, for $\omega \lesssim v_F /L_0$, we will argue that the DOS is 
given by the expression in Eq. (\ref{dos0}).

In Sec. II, we will consider a system of spinless electrons
with LRCI and disorder. After bosonizing the Hamiltonian, we well use the
RG equations to obtain the localization length $L_0$ in terms of the 
disorder strength. We will also present a physical understanding of the 
expression for $L_0$. In Sec. III, we will obtain expressions for the DOS for 
different ranges of values of $\omega$. In Sec. IV, we will briefly discuss 
the case of a screened Coulomb interaction.

\section{Bosonization and the localization length}

The Hamiltonian of a disordered system of spinless electrons with LRCI 
consists of three parts,
\be 
H =H_0 + H_C + H_{\rm random} ~.
\ee
The noninteracting part $H_0$ and the Coulomb interaction part $H_C$ are 
written in terms of the continuum chiral electron fields as follows,
\ba
H_0 &=& v_F \int_{-\infty}^\infty dx \left( -i \psi_R^{\dagger} \partial_x
\psi_R + i \psi_L^{\dagger} \partial_x \psi_L \right) ~, \\
H_C &=& \frac{1}{2} ~\int_{-\infty}^\infty dx dy ~U(x-y) \rho(x) \rho(y) ~,
\label{hoc}
\ea
where the form of $U(x-y)$ will be specified later.
The fields $\psi_R(\psi_L)$ are the right-moving (left-moving) electron 
operators; they are related to the lattice electron operators
by
\be
c(x) = \sqrt a \left[ e^{ik_Fx} \psi_R(x) + e^{-ik_Fx} \psi_L(x) \right] ~,
\ee
where $a$ is the lattice spacing. The density $\rho = \rho_R + \rho_L$, where
$\rho_R$ ($\rho_L$) is the normal-ordered density of the right (left) moving 
electrons.

In our model, the random part of the Hamiltonian consists of two
parts in the continuum limit, (i) the forward scattering part, $H_f$, where 
the scattered electrons remain in the vicinity of the same Fermi point, 
and (ii) the backward scattering part, $H_b$, where an electron is 
scattered from $-k_F$ to $k_F$, or vice versa. 

Using the rules of bosonization \ct{sen,gogolin},
we rewrite the full Hamiltonian in the bosonic language. The low-energy
and long-wavelength excitations of the noninteracting part can be written as
\be
H_0=\frac{v_F}{2 \pi} \int_{-\infty}^\infty dx \left[ (\partial_x 
\theta(x))^2+ (\partial_x \phi (x))^2 \right] ~,
\label{ho}
\ee
while the Coulomb part is given by
\be
H_C =\frac{1}{2\pi^2} ~\int_{-\infty}^\infty dx dy ~U(x-y) \partial_x 
\phi(x) \partial_y \phi(y) ~.
\label{hc}
\ee
The forward scattering part is given by
\ba
H_f &=&\int_{-\infty}^\infty dx ~h(x) \left[ \psi_R^{\dagger}(x)\psi_R(x)+
\psi_L^{\dagger}(x)\psi_L(x) \right] \non \\
&=&-\frac{1}{\pi} \int_{-\infty}^\infty dx ~h(x) \partial_x \phi (x) ~.
\ea
The forward scattering is due to the real random field $h(x)$, and its effect 
can be taken care of by a rescaling of the bosonic field $\phi$ 
\ct{giamarchi}; henceforth, we shall ignore the forward scattering part.
The back-scattering part can be written as
\ba
H_b&=& \int_{-\infty}^\infty dx ~\left[ \xi (x) \psi_R^{\dagger}(x)
\psi_L(x)+\xi^*(x) \psi_L^{\dagger}(x)\psi_R(x) \right] \non \\ 
&=&\frac{1}{\pi \alpha} \int_{-\infty}^\infty dx ~\left[ \xi (x)
e^{i(2 \phi (x) + 2 k_F x)} + \textrm{h.c.} \right] ~.
\label{hb}
\ea
where h.c. denotes the hermitian conjugate. The back-scattering is due to 
a complex random field $\xi(x)$ with the probability distribution
\be
P [\xi (x)] = \exp \left[ -D_{\xi}^{-1} \int dx \xi^* (x) \xi(x) \right] ~,
\label{pxi}
\ee
so that $< \xi (x) > = 0$ and $< \xi^* (x) \xi (x')> = D_\xi \delta (x-x')$.

The Hamiltonians given in Eqs. (\ref{ho}), (\ref{hc}) and (\ref{hb}), along 
with the probability distribution (\ref{pxi}) of the random field $\xi$, 
constitute the complete low-energy and long-wavelength description of the 
model in the bosonic language. We now rewrite the Coulomb interaction in Eq. 
(\ref{hc}) in momentum space as
\ba
& & \frac{1}{2\pi^2} \int dx dy U(x-y) \partial_x \phi (x) \partial_y
\phi(y) \non \\
&=& \frac{1}{4\pi^3} \int dk k^2 \hat{U} (k) \hat{\phi}(k) \hat{\phi} (-k) ~,
\label{u0}
\ea
where $\hat{U}(k)$ is the Fourier transform of the Coulomb interaction. We 
can expand $\hat U(k)$ in powers of $k$ as
\ba
\hat{U}(k)=\hat{U}_0+\hat{U}_1 k +\hat{U}_2 k^2+.... ~.
\label{vk}
\ea
{}From the unperturbed Hamiltonian in Eq. (\ref{ho}), we find the naive 
scaling dimensions of various quantities to be
\ba
[\omega] ~=~ [k] ~=~ ({\rm Length})^{-1} ~, \quad [{\hat \phi}] ~=~ 
({\rm Length})^1 ~.
\ea
Considering the scaling dimensions of the various terms in Eq. (\ref{vk}), we 
conclude that only the constant term is marginal under renormalization, whereas
the terms involving powers of $k$ are irrelevant. Hence, the important 
contribution arising from the Coulomb interaction term in Eq. (\ref{u0}) can be
written as
\ba
H_C &=& \frac{\hat{U}_0}{4\pi^3} \int dk k^2 \hat{\phi}(k) \hat{\phi} (-k) 
\non \\
&=& \frac{\hat{U}_0}{2\pi^2} \int dx \left( \partial_x \phi (x) 
\right)^2 ~.
\ea
The constant $\hat{U}_0$ is given by
\ba
\hat{U}_0=\int_{-L/2}^{L/2} dx ~\frac{q^2}{\sqrt{x^2 + d^2}} = 
2q^2 \ln {\frac{L}{d}} ~,
\label{hc1}
\ea
where $q$ is the charge of the electron, and $d$ is the width of the wire 
which is of the order of the lattice spacing $a$; it ensures convergence of 
the integral in Eq. (\ref{hc1}) near $x=0$. 
The meaning of the cut-off length $L$ will be discussed below. (We have 
assumed that $L >> a,d$; this will be justified later for various cases of 
interest). We now define a dimensionless quantity
\ba
\eta = \frac{2q^2}{\pi v_F} \ln{\frac{L}{a}} ~.
\label{eta}
\ea
Then the Hamiltonian can be written in the form $H= H_0 + H_C + H_b$, where
\ba
H_0 + H_C &=& \frac{v_F}{2\pi} \int dx \left[ (\partial_x \theta (x))^2
+(1+\eta)(\partial_x \phi (x))^2 \right] ~. \non \\
& &
\label{hfull1}
\ea
Eq. (\ref{hfull1}) can be brought into a standard form by defining
\ba
K=\frac{1}{\sqrt{1+\eta}} ~, \quad {\rm and} \quad u=v_F \sqrt{1+\eta} ~.
\label{ku}
\ea
We then arrive at the expression
\ba
H_0 + H_C &=& \frac{u}{2 \pi} \int d x \left[ K (\partial_x \theta
(x))^2+\frac{1}{K} (\partial_x \phi (x))^2 \right] ~. \non \\
& &
\label{hfull2}
\ea

In the imaginary time representation, the quadratic and random parts of the 
action are respectively given by
\ba
S_0 &+& S_C \non \\
&=& \frac{u}{2\pi} \int dx d\tau \left[ K (\partial_{\tau}
\phi (x,\tau) )^2 + \frac{1}{K} (\partial_x \phi (x,\tau))^2 \right], \non \\
S_b &=& \frac{1}{\pi \alpha} \int dx d\tau \left[ \xi (x) e^{i(2 \phi 
(x,\tau) +2 k_F x)} + h.c. \right].
\ea
The randomness in the back-scattering term is dealt with using the standard 
replica method and averaging over the randomness using the distribution in Eq. 
(\ref{pxi}). The final form of the disorder-averaged $n$-replicated action is 
found to be \ct{giamarchi}
\ba
&S_n& \non \\
&=&\frac{u}{2\pi} \sum_a \int dx d\tau \left[ K (\partial_{\tau} \phi^a 
(x,\tau))^2 + \frac{1}{K} (\partial_x \phi^a (x,\tau))^2 \right] \non \\
&-&\frac{D_\xi}{(\pi \alpha)^2} \sum_{a,b} \int dx d\tau d\tau' \cos {\left( 2
\phi^a (x,\tau) - 2 \phi^b (x,\tau') \right)}, \non \\
& &
\label{srep}
\ea
where $a,b$ are the replica indices running from $1$ to $n$. Eq. (\ref{srep}) 
is the same as the replicated action of a one-dimensional disordered electronic
system given in Ref. \ct{giamarchi}. However, the back-scattering term 
arising due to the spin of the electrons is absent in the present case of 
spinless electrons. The RG equations of the different parameters are given by 
\ct{giamarchi}
\ba
\frac{dD}{dl} &=& (3-2K) D ~, \non \\ 
\frac{dK}{dl} &=& -\frac{1}{2} K^2 D ~, \non \\
\frac{du}{dl} &=& -\frac{u K}{2}D ~.
\label{rg}
\ea
where $D = (2D_\xi \alpha)/\pi u^2$, and $l=\ln (L/a)$ is the logarithm of 
the length scale. The RG equations given above show a quantum phase transition
at $K = 3/2$. For $K > 3/2$, the disorder is irrelevant. For $K < 3/2$, the 
disorder grows under RG and leads to localization. Thus the system
undergoes a zero-temperature localized-delocalized transition at $K = 3/2$.

The localization length, denoted by $L_0$, 
can be obtained by integrating the RG equations in (\ref{rg}). At the 
microsocopic length scale $a$, we begin with some values of $D$ (much smaller 
than 1), and $u, ~K$ given in Eq. (\ref{ku}). (The latter requires a 
knowledge of the length scale $L$ through Eq. (\ref{eta}); this will be 
determined self-consistently by the solution of the RG equations). Since we 
are assuming that $K << 1$, we find (as can be verified numerically using Eqs. 
(\ref{rg})) that the quantities $K$ and $u$ flow very little while $D$ flows 
from a small number of order $D_\xi$ to a number of order 1. Thus the 
length scale at which the disorder strength becomes of order 1 is given by 
\be
\frac{L_0}{a} \sim \left( \frac{1}{D_\xi} \right)^{1/(3 - 2K)} ~.
\label{lo}
\ee
We identify this as the localization length whose physical meaning will be
discussed below.

In the limit of weak disorder ($D_\xi \to 0$), the form of the localization 
length in Eq. (\ref{lo}) implies, in a self-consistent manner, that (i) 
$L_0 >>a$, (ii) the parameter $\eta >> 1$ (replacing $L$ by 
$L_0$ in Eq. (\ref{eta})), and (iii) the interaction parameter $K << 1$ due to
Eq. (\ref{ku}). Hence the localization length in the $K \to 0$ limit assumes 
the classical value
\be
\frac{L_0}{a} \sim \left( \frac{1}{D_\xi} \right)^{\frac{1}{3}} ~.
\ee

[Note that the bosonization approach cannot be used in the extremely 
localized limit ($L_0 \sim a$) studied in Refs. \ct{raich,mobius,vojta}. In 
that limit, one cannot use a continuum description, and thus a bosonized 
description of the system is not possible].

The significance of the localization length is as follows. Although two 
electrons which are separated by more than the distance $L_0$ do interact with
each other through the Coulomb potential, the overlap of their wave functions 
is exponentially small, and hence their positions are uncorrelated. Such 
interactions will therefore only contribute to a uniform and static chemical 
potential which is the same for all electrons. Namely, the Coulomb potential 
felt by an electron from other electrons which are separated from it by a 
distance larger than $L_0$ is described by a part of Eq. (\ref{hoc}) given by
\ba
H_C = \frac{1}{2} \int_{-\infty}^\infty dx ~\rho(x) ~\rho_0 ~[ & &
\int_{x+L_0}^\infty dy ~U(x-y) \non \\
& & +~ \int_{-\infty}^{x-L_0} dy ~U(x-y) ~],
\label{hcu}
\ea
where $\rho_0$ is the average density. Eq. (\ref{hcu}) represents a 
uniform one-body potential.
On the other hand, when two electrons are closer to each other than $L_0$ 
that their positions are correlated; then their Coulomb interaction has to be 
included in the quadratic part of the bosonic Hamiltonian (Eq. (\ref{hc}) to 
be specific) which governs the density fluctuations.
In short, portions of the the system which are separated by distances larger 
than $L_0$ are uncorrelated, whereas, within a distance $L_0$, the system can 
be described in terms of a TLL with $K << 1$.

Before ending this section, we would like to present a simple 
physical understanding of the important result in Eq. (\ref{lo})
which does not make use of the replica idea. First, let us consider a weak 
$\delta$-function impurity of strength $V_0$ at one point in a TLL. According 
to an RG equation derived by Kane and Fisher \ct{kane}, the impurity strength
$V$ flows according to the equation $dV/dl = (1-K)V$; hence, at a distance 
scale L, we have $V(L) \sim V_0 ~(L/a)^{1-K}$. From quantum mechanics, we know 
that the reflection amplitude for an electron scattering from a single impurity
is proportional to $V$ if $V/v_F$ is small. Now suppose that each site in the 
lattice has a random impurity of strength $V(x)$ which 
satisfies $<V(x) V(y)> = D_\xi \delta (x-y)$. This means that at each site,
$V(x)$ is of the order of $\sqrt{D_\xi}$, and its sign is random. Over a length
$L$, there are $L/a$ reflections since there is a reflection at each site. In 
order that these $L/a$ reflections should add up to a total reflection 
amplitude of order 1 (which would localize the electron), we must have 
$\sqrt{L/a} ~V(L) \sim 1$ (assuming that $L/a$ is large, and using the 
well-known result for the sum of a large number of random terms in a 
Brownian motion). This gives us the estimate $L_0/a \sim 1/(V(L_0))^2$, i.e.,
$L_0/a \sim 1/(V_0 (L_0/a)^{1-K})^2$. Since $V_0 \sim \sqrt{D_\xi}$, we
obtain $L_0/a \sim (1/D_\xi)^{1/(3-2K)}$].

Our approach shows that the two ways of introducing disorder, (i) at
random positions with a density $n_i$ but with equal strengths $V_0$ as in Ref.
\ct{lee}, and (ii) at all sites but with random strengths $\xi (x)$ as in the 
present case, have the same effect. This is because the only thing 
which matters finally is the disorder parameter $D_\xi$ which appears in the 
expression in Eq. (\ref{lo}) for the localization length \ct{giamarchi}. The 
exact nature of the randomness is not relevant for determining the 
form of $L_0$. The relation between the two sets of disorder parameters
is given by $D_\xi \sim V_0^2 n_i$. Using this relation, one can write the 
expression for $L_0$ given in Ref. \ct{lee} directly in terms of $D_\xi$ and 
check that $L_0$ is proportional to $(1/D_\xi)^{1/3}$ (except for a logarithmic
correction arising due to the LRCI \ct{maurey}). This is also what the RG 
equation yields for $\eta >> 1$. Hence, our approach provides an 
independent (RG) derivation of $L_0$ which is an alternative to the derivation
using considerations of energy minimization \ct{maurey,fukuyama}. We have
also presented above an alternative understanding of the expression in
Eq. (\ref{lo}) which does not use the replica idea and is based only on the
RG equation for a single impurity.

\section{Density of states}

We are now ready to discuss the density of states. For energies satisfying 
$\omega \gtrsim v_F /L_0$, the disorder parameter $D$ is small, and our 
system can be described by a clean TLL. The DOS of a TLL is given by the 
relation 
\ct{sen} 
\ba
D (\omega) & \sim & |\omega|^{\beta} ~, \non \\
\beta &=& \frac{(1-K)^2}{2 K} = \frac{{\sqrt {1+\eta}}}{2} + 
\frac{1}{2{\sqrt {1+\eta}}} - 1 ~,
\label{dos3}
\ea
where we have used the relation $K = 1/\sqrt{1+ \eta}$. This is the result 
quoted in Eq. (\ref{dos2}). However, we now have to determine what value of 
$L$ one should take in Eq. (\ref{eta}) in order to determine the value of 
$\eta$ to be used in Eq. (\ref{dos3}).

It turns out that the value of $\omega$ itself determines the appropriate 
value of $\eta$ to use in Eq. (\ref{dos3}). If $\omega = v_F /L_0$, then the
length scale of interest is $v_F /\omega$ which is equal to $L_0$. The
value of $\eta$ to use in Eq. (\ref{dos3}) is then given self-consistently by 
Eqs. (\ref{eta}) (with $L$ replaced by $L_0$), (\ref{ku}) and (\ref{lo}).
However, if $\omega > v_F /L_0$, then the length scale of interest is $L
= v_F /\omega$ which is smaller than $L_0$. Then the Coulomb interaction
should be cut-off at the length scale $L$ as indicated in Eq. 
(\ref{hc1}). The reason for this lies in the basic idea behind the
RG method, namely, that the properties of a system at a length scale $L$ 
(or, equivalently, at an energy scale $v_F /L$) are governed essentially 
by the modes whose wavelengths are smaller than $L$. (The modes whose
wavelengths are larger than $L$ only contribute to a uniform and static
chemical potential in the sense described in Eq. (\ref{hcu})).
If the Coulomb interaction is cut-off
at the distance $L$, the value of $\eta$ to be used in Eq. (\ref{dos3}) is
simply given by Eq. (\ref{eta}). We thus conclude that as long
as $\omega \gtrsim v_F /L_0$, the DOS of states is given by Eq. (\ref{dos3})
where $\eta$ depends on $\omega$ through Eq. (\ref{eta}) with $L =v_F /\omega$.

We note that the expression in Eq. (\ref{dos2}) agrees with the semiclassical 
result in Eq. (\ref{dos1}) in the limit $\eta >> 1$.
Moreover, Eq. (\ref{dos2}) correctly reproduces the expression of 
the DOS of a noninteracting disordered electronic system (for which $\eta =
0$ and $K=1$), where there is a finite DOS at the Fermi energy \ct{halperin}.

We now consider the case of low energies satisfying $\omega < v_F /L_0$.
In this case, the RG equations in (\ref{rg}) imply that the disorder is
strong; hence the system cannot be described by a clean TLL. However, this
is precisely the regime described by the localized limit discussed in
Refs. \ct{raich,mobius,vojta}, where the localization length $L_0$ is smaller
than the length scale of interest, namely, $L= v_F /\omega$. We therefore
expect the DOS in this regime to be described by Eq. (\ref{dos0}).

We have thus found expressions for the DOS in the two regimes $\omega \gtrsim
v_F /L_0$ and $\omega \lesssim v_F /L_0$ (strictly speaking, $\omega >> v_F 
/L_0$ and $\omega << v_F /L_0$ respectively). The DOS must of course cross
over smoothly from one regime to the other, but this cross over is difficult
to determine analytically.

Let us summarize our key results. In the limit of weak disorder, the 
system is power-law correlated only
over a large distance of order $L_0$, called the localization or pinning 
length. At that length scale, the system is described by a TLL
in which the LRCI drastically reduces the interaction parameter $K$ to a 
value much smaller than unity; such a value of $K$ is quite uncommon in the 
literature. (Thus the Coulomb interaction has the same effect as a very strong
short-range interaction)! We note that in the absence of disorder, a system 
with a LRCI is {\it not} equivalent to one with a short-range interaction. 
For a clean system, the LRCI leads to a Wigner crystal \ct{schulz} for which 
a TLL description is not valid. In our formalism, we can see this by noting 
that if the disorder was absent (i.e., $D_\xi =0$), $L_0$ 
would be infinite; then $\eta$ and $K$ would respectively be $\infty$ and
zero, and the TLL description would no longer be valid. On the other hand, 
for the disordered but noninteracting system (i.e., $q^2 =0$), we have 
$\eta =0$ and $K =1$ even if $L_0$ is very large.

The TLL description enables us 
to derive the DOS easily using well-known relations without having to make 
use of a semiclassical approach \ct{lee}. Even though the final expressions 
for the DOS in Eqs. (\ref{dos1}) and (\ref{dos2}) match in the limit of 
$\eta >>1$, the expression in Eq. (\ref{dos1}) derived in Ref. \ct{lee}
shows a gap at the Fermi energy of the form $\omega^{1/2}$, and thus fails to 
reproduce the result for the noninteracting system. Our TLL approach correctly
captures the physics of both the interacting and noninteracting limits by 
using the appropriate values of the interaction parameter $K$. 

Finally, beyond the length scale $L_0$, the TLL description is no longer
valid since the disorder strength is of order 1. In this regime, the DOS
is given by Eq. (\ref{dos0}); thus as $\omega \rightarrow 0$, the DOS goes to
zero logarithmically rather than as a power-law.

\section{Discussion}

It is interesting to consider what would happen if the Coulomb interaction
was screened. We then have
\ba
\hat{U}_0 &=& \int_{-L/2}^{L/2} dx ~\frac{q^2 e^{-x/L_s}}{\sqrt{x^2 + d^2}} ~,
\non \\
\eta &=& \frac{\hat{U}_0}{\pi v_F} ~, \quad {\rm and} \quad K ~=~ 
\frac{1}{\sqrt{1+\eta}} ~,
\label{hc2}
\ea
where $L_s$ is the screening length. It is clear that $\hat{U}_0$ now has a 
finite limit, and that $\eta$ and $K$ no longer approach $\infty$ 
and zero respectively, as $L \rightarrow \infty$. For the case of weak 
disorder, the localization length $L_0$ is again given by the expression in 
Eq. (\ref{lo}), where $K$ is self-consistently determined by Eq. 
(\ref{hc2}) with $L$ replaced by $L_0$.

Let us consider the case in which the screening length $L_s$ is much smaller 
than $L_0$. Now there are three ranges of energy to consider for the DOS, 
(i) $\omega \gtrsim v_F /L_s$, (ii) $v_F /L_s \gtrsim \omega \gtrsim
v_F /L_0$, and (iii) $v_F /L_0 \gtrsim \omega$. In regimes (i) and (ii), 
arguments similar to those presented in Sec. III will show that the system can
again be described by a clean TLL in which the DOS is given by Eq. 
(\ref{dos3}), with $\eta$ being determined by $\omega$ through Eq. 
(\ref{hc2}) with $L = v_F /\omega$. Clearly, $\eta$ will increase as $\omega$ 
decreases in regime (i). However, $\eta$ will stop changing appreciably once
$L$ becomes much bigger than $L_s$; hence, $\eta$ will be almost constant in
regime (ii). Finally, in regime (iii), the disorder is of order 1, the
system is in the localized limit (the length scale of interest is
larger than $L_0$), and the DOS is given by Eq. (\ref{dos0}).

Before ending, we would like to mention that the results obtained here 
for the localization length and the DOS can be used to study 
disordered antiferromagnetic spin chains. In particular, the spin-1/2 chain 
with random magnetic field and random exchanges of various kinds has been 
studied extensively in recent years \ct{doty,fisher,bunder}. Our analysis can
be applied to such systems after mapping the spins to Jordan-Wigner fermions
\ct{doty}.

\section*{Acknowledgments}

AD and LF gratefully acknowledge important and encouraging discussions with R.
Oppermann. AD also acknowledges the hospitality of the Institut f\"ur 
Theoretische Physik und Astrophysik, Universit\"at W\"urzburg, where the 
initial part of the work was done, and Deutsche Forschunggemeinschaft for 
financial support via project OP28/5-2. DS thanks the Department of Science 
and Technology, India for financial support through Grant No. SP/S2/M-11/00.

\end{multicols}

\end{document}